# DEVELOPMENT AND MAINTENANCE OF XML-BASED VERSUS HTML-BASED WEBSITES: A CASE STUDY

Mustafa Atay
*Department of Computer Science*
*Winston-Salem State University*
*Winston-Salem, NC 27110 USA*
*ataymu@wssu.edu*

**ABSTRACT**

HTML (Hyper Text Markup Language) has been the primary tool for designing and developing web pages over the years. Content and formatting information are placed together in an HTML document. XML (Extensible Markup Language) is a markup language for documents containing semi-structured and structured information. XML separates formatting from the content. While websites designed in HTML require the formatting information to be included along with the new content when an update occurs, XML does not require adding format information as this information is separately included in XML stylesheets (XSL) and need not be updated when new content is added. On the other hand, XML makes use of extra tags in the XML and XSL files which increase the space usage. In this study, we design and implement two experimental websites using HTML and XML respectively. We incrementally update both websites with the same data and record the change in the size of code for both HTML and XML editions to evaluate the space efficiency of these websites.

**KEYWORDS**

XML, XSLT, HTML, Website Development, Space Efficiency

## 1. INTRODUCTION

HTML (Hyper Text Markup Language) [1] has been the primary tool for designing and developing web pages for a long time. HTML is primarily intended for easily designing and formatting web pages for end users. Web pages developed in HTML are unstructured by its nature. Therefore, programmatic update and maintenance of HTML web pages is not an easy task.

XML (Extensible Markup Language) [2] is a markup language for documents containing semi-structured and structured information. Structured information has uniform format and contains both content and some indication of what role that content plays within the context of a document. XML utilizes the efficient use of richly structured documents over the web.

Unlike HTML, XML comes with a family of technologies such as XSLT (XML Stylesheet Language Transformations) [3], which is used to style XML documents, XPath (XML Path Language) [4], which is used to locate nodes in an XML document, and XML Schema [5], which is used to define content rules for XML documents. With its semi-structured nature and accompanying technologies, XML enables programmatic updates and maintenance of web pages.

While content and formatting information are placed together in an HTML document, XML separates formatting from the content. Websites designed in HTML require formatting information to be included along with the new content when an update occurs. XML does not require adding format information when an update occurs as this information is separately included in XML stylesheets (XSL) [6] and need not be updated when new content is added. On the other hand, XML makes use of extra tags in the XML and XSL files which increase the space usage.

XML is suggested as a new website architecture by SUNY Center for Technology in Government (CTG) [7]. A number of potential benefits of migrating a website from HTML to XML are listed in a technical paper from SUNY CTG [8]. Among those benefits, CTG did not report the space efficiency of an XML website which is one of the key issues for enhanced performance and cost savings. In this paper, we elaborate and address this issue and focus on the comparison of space efficiency between an HTML-based website and its corresponding XML-based edition. We also show that XSLT enables programmatic updates of an XML-based website.

The rest of the paper is organized as follows: Section 2 briefly describes the website for case study. The process of migrating an HTML website to an XML website is explained in Section 3. Section 4 presents the results of the experimental study. Finally, Section 5 concludes the paper and points out some future work.

## 2. CASE STUDY

We design two experimental websites, one in HTML and the other in XML, for the Summer Undergraduate Research Experience (SURE) Program in Science, Technology, Engineering and Mathematics for the purpose of this research study. An undergraduate summer scholar contributed to the development and maintenance of these websites for six weeks in summer 2012. Therefore, this study also contributed to the encouragement and involvement of undergraduate students in research work.

We designed a plain and informative website rather than a fancy one for the SURE Program as our testbed. The website contains a Home page, an Application page, a Personnel page and pages for Participants from 2009 thru 2012. A snapshot of the testbed website is shown in Figure 1.

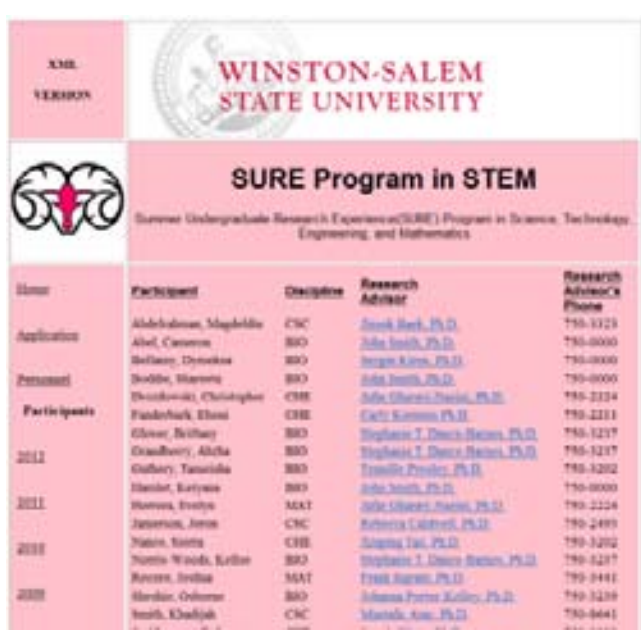

Figure 1. A snapshot of the testbed website

```
13  <tr bgcolor="FF0000" valign="top">
14    <td height="79" bgcolor="FF0000"><p class="style1"><br></br>  <a href="home.html" style="color: #FFFFFF"/>Home</a
15    <p></p>
16          <a href="application.html" style="color: #FFFFFF"/>Application</a><br />
17        <p></p>
18          <a href="personnel.html" style="color: #FFFFFF"/>Personnel</a><br />
19    <p></p>
20           <strong>Participants</strong>
21    <br> <class="style1">  <a href="2012.html" style="color: #FFFFFF"/>2012</a>
22    <p></p>
23      <a href="2011.html" style="color: #FFFFFF"/>2011</a>
24    <p></p>
25      <a href="2010.html" style="color: #FFFFFF"/>2010</a>
26    <p></p>
27      <a href="2009.html" style="color: #FFFFFF"/>2009</a></br>
28
29   <td height="600">
30   <table border="0" width="600">
31   <tr>
32   <th align="left" width="150"><font face="Arial" color="white"><u>Participant</u></font></th>
33  <p><th align="left" width="85"><font face="Arial" color="white"><u>Discipline</u></font></th></p>
34  <p><th align="left" width="217"><font face="Arial" color="white"><u>Research Advisor</u></font></th></p>
35  <p><th align="left" width="0"><font face="Arial" color="white"><u>Advisor's Phone</u></font></th></p>
36  </tr>
37  <tr>
```

Figure 2. HTML code segment for Participants page

While the data in Home, Application and Personnel pages are either fixed or rarely updated, Participants pages are updated on a regular basis as new participants join the SURE Program. Since we want to show the difference in maintaining both HTML and XML editions of the testbed site, Participants pages played an important role in our experimental study.

## 3. MIGRATING FROM HTML TO XML

We make use of HTML, XML, XSLT and XPath technologies in this study. We design and implement two experimental websites for SURE Program using HTML and XML respectively. We ensure that these websites are identical.

Developing HTML edition of the website was straight forward and hassle-free. A snapshot of the HTML markup for Participants page is shown in Figure 2. As seen in the Figure, the page content is mixed with the page format in HTML. For instance, at line 27 in Figure 2, a hyperlink is introduced and styled for the 2009 participant list. At line 32 of Figure 2, the column header *Participant* is displayed as left aligned with Arial font type and white color.

While designing and developing XML edition of the website, we separated content of the pages from their format. While content goes to the XML files, style information is placed into the XSLT files. We make use of XPath in the XSLT code to locate the XML nodes for styling.

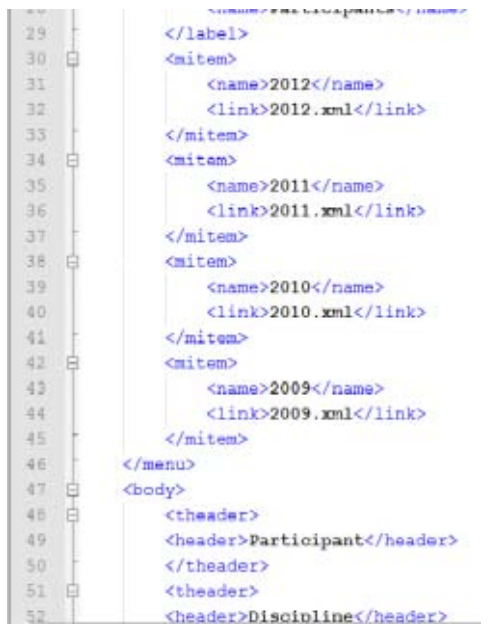

```
29         </label>
30         <mitem>
31             <name>2012</name>
32             <link>2012.xml</link>
33         </mitem>
34         <mitem>
35             <name>2011</name>
36             <link>2011.xml</link>
37         </mitem>
38         <mitem>
39             <name>2010</name>
40             <link>2010.xml</link>
41         </mitem>
42         <mitem>
43             <name>2009</name>
44             <link>2009.xml</link>
45         </mitem>
46     </menu>
47     <body>
48         <theader>
49         <header>Participant</header>
50         </theader>
51         <theader>
52         <header>Discipline</header>
```

Figure 3. XML code segment for Participants page

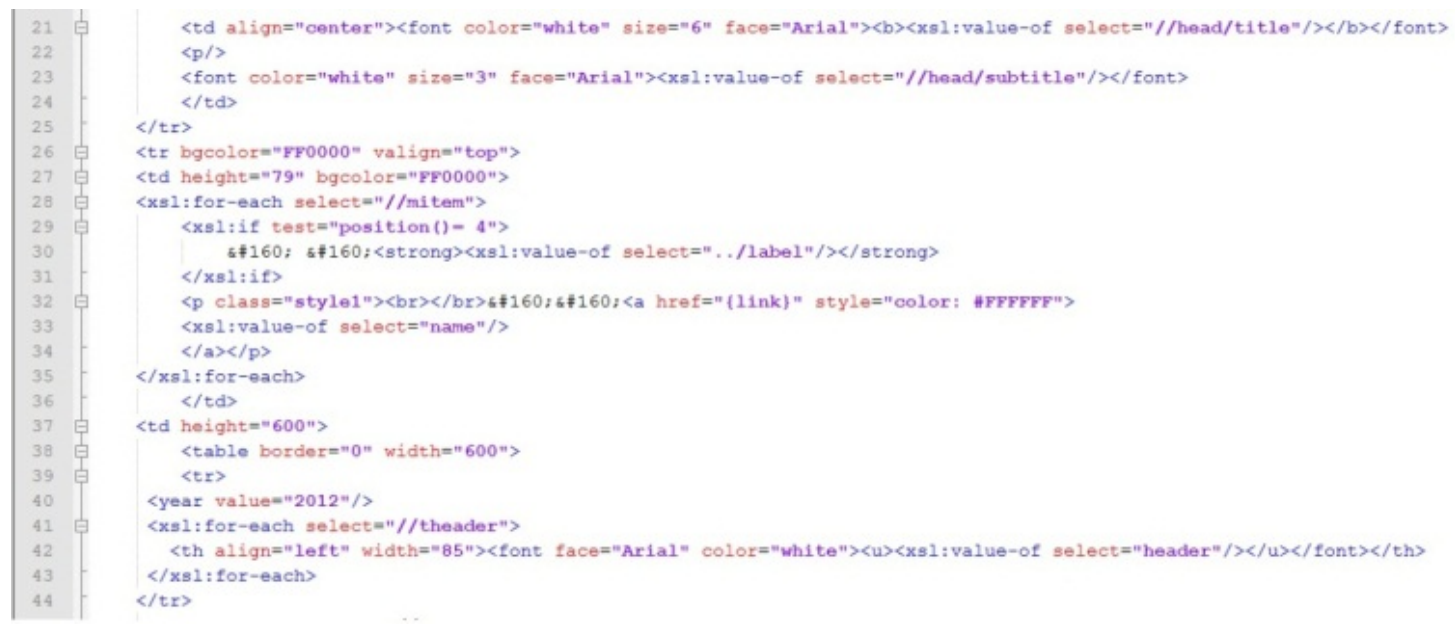

```
21         <td align="center"><font color="white" size="6" face="Arial"><b><xsl:value-of select="//head/title"/></b></font>
22         <p/>
23         <font color="white" size="3" face="Arial"><xsl:value-of select="//head/subtitle"/></font>
24         </td>
25     </tr>
26     <tr bgcolor="FF0000" valign="top">
27     <td height="79" bgcolor="FF0000">
28     <xsl:for-each select="//mitem">
29         <xsl:if test="position()= 4">
30                <strong><xsl:value-of select="../label"/></strong>
31         </xsl:if>
32         <p class="style1"><br></br>  <a href="{link}" style="color: #FFFFFF">
33         <xsl:value-of select="name"/>
34         </a></p>
35     </xsl:for-each>
36         </td>
37     <td height="600">
38         <table border="0" width="600">
39         <tr>
40      <year value="2012"/>
41      <xsl:for-each select="//theader">
42        <th align="left" width="85"><font face="Arial" color="white"><u><xsl:value-of select="header"/></u></font></th>
43      </xsl:for-each>
44     </tr>
```

Figure 4. XSLT code segment for Participants page

There are a couple of design choices for the XML documents. It is important to design the XML files in a way that will minimize the XSLT code for styling. The structure and labeling of XML file should reflect the semantics of the original HTML file. Therefore, document authors should spend some quality time while deciding on the design of XML files.

A snapshot of the XML markup for Participants is shown in Figure 3. Line 42-45 shows the content for 2009 participants link while the style information is not included. Line 49 shows the column header *Participant* without its format information.

We coded the style information of the underlying HTML pages in XSLT programs. XSLT uses XPath to locate the XML nodes for formatting. XSLT is like a high level programming language. It includes variables, selection and looping constructs. During the process of migrating an HTML page to its XML equivalent, most of the time is spent for coding XSLT programs.

A snapshot of the XSLT code which styles the XML document for Participants is shown in Figure 4. The for loop at lines 28-35 styles all the navigation menu items including the participants link for year 2009. The for loop at lines 41-43 formats the all the column headers including the header *Participant.*

The *//mitem* expression at line 28 is an XPath expression which locates all the *mitem* elements in the underlying XML documents. Once the context node is set as *mitem*, the *../label* expression selects the *label* element which is the preceding sibling of the fourth *mitem* element.

## 4. EXPERIMENTAL STUDY

In the experimental study, we update both websites over a period of time to evaluate and compare space efficiency of both HTML-based and XML-based websites. We choose number of characters as our space measurement metric. Prior to recording our findings, we normalize all HTML, XML and XSLT files by removing the unnecessary white spaces. We first record the size of initial code to create both of these websites. Then, we incrementally update both websites with the same data and record the change in the size of code for both HTML and XML editions.

We only added new Participants list while updating the websites. We did not make any changes in Home, Application and Personnel pages during the update process. While the chart shown in Figure 5 compares the total number of characters for all the files in both HTML and XML editions, the chart given in Figure 6 only compares the total number of characters in Participants pages.

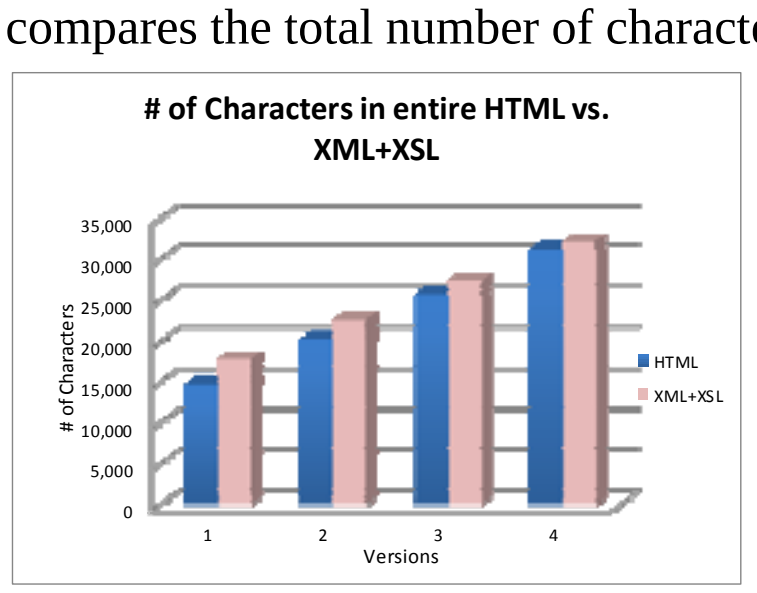

Figure 5. Total # of characters in HTML vs. in XML+XSL

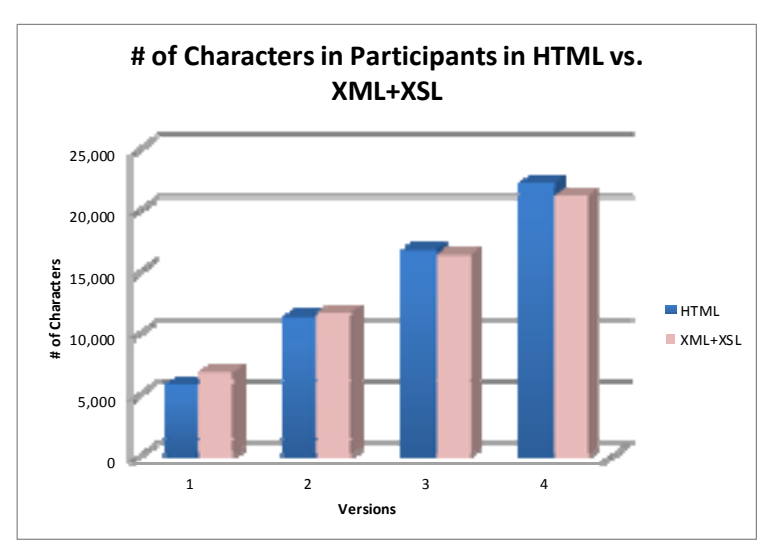

Figure 6. Total # of characters of Participants in HTML vs. in XML+XSL

We observed from Figure 5 that initially HTML uses less space than XML+XSL combination. As both editions are incrementally updated, although the space usage gap between HTML and XML editions is getting closed, HTML slightly outperforms the XML edition. Although the updated pages reduce the amount of space used for XML, static pages in the calculation favor for the HTML edition.

In Figure 6, we only measured the space efficiency of the dynamic parts of the testbed websites which only include the Participants files. The chart in Figure 6 shows that as the websites are updated, the space usage gap for the updated pages is closed and XML edition outperformed the HTML edition.

## 5. CONCLUSIONS

Experimental study showed that XML is more space efficient than HTML especially for pages with structured and semi-structured data. XML brings overhead for pages with unstructured data. HTML would be the choice for space efficiency of websites with unstructured data. HTML initially consumes less space than XML as XML and XSLT introduces auxiliary tags which bring overhead. If a website contains dynamic pages which are updated often, XML outperforms HTML in the long run.

XSLT utilized partial programmatic updates of the XML website while the entire HTML website is manually updated. While XSLT enabled the programmatic format of the newly added content, the new raw content itself is introduced to XML data files with the intervention of a document author.

We will consider presenting website content in different output formats such as printer friendly, PDF, etc. and see how it will affect the space efficiency of the XML-based versus HTML-based websites in the future work.

This study also contributed to the undergraduate research as an undergraduate scholar involved effectively in the development of experimental HTML and XML testbed websites. We will encourage participation of undergraduate scholars in the future extensions of this research work as well.

## ACKNOWLEDGEMENT

This study is partially funded by NSF HBCU-UP Implementation Grant #0927905. Many thanks for the contributions of the undergraduate scholar Khadijah Smith.